\documentclass[twocolumn,showpacs,superscriptaddress,prb,aps,floatfix]{revtex4}

\usepackage{graphicx}
\usepackage{rotating}
\usepackage{amsmath}
\usepackage{soul}
\usepackage[usenames,dvipsnames]{color}

\begin{document}

\title{Thickness dependent structural and electronic properties of $\rm CuO$
adsorbed on SrTiO$_3$(100):  a hybrid density functional theory study}

\author{C. Franchini}
\affiliation{
Faculty of Physics, Universit\&quot;at Wien and Center for Computational Materials Science,  A-1090 Wien, Austria}

\author{Xing-Qiu Chen}
\affiliation{
Shenyang National Laboratory for Materials Science,
Institute of Metal Research, Chinese Academy of Sciences,
72 Wenhua Road, Shenyang 110016, China
}
\author{R. Podloucky}
\affiliation{Institute for Physical Chemistry, Universit\&quot;at Wien,
Sensengasse 8/7, A-1090 Wien, Austria.}

\date{\today}

\begin{abstract}
We discuss the structural and electronic properties of tetragonal CuO grown on $\rm SrTiO_3$(100)
by means of hybrid density functional theory. Our analysis explains the anomalously large Cu-O vertical 
distance observed in the experiments ($\approx$2.7\AA) in terms of a peculiar frustration between 
two competing local Cu-O environments characterized by different in-plane and out-of-plane
bond lengths and Cu electronic populations.
The proper inclusion of substrate effects is crucial to understand the 
tetragonal expansion and to reproduce correctly the measured valence band spectrum for a CuO thickness 
of 3-3.5 unit cells, in agreement with the experimentally estimated thickness.
\end{abstract}

\maketitle

\section{Introduction}

Very recently, Siemons \emph{et al.} \cite{siemons} synthesized  CuO with a
tetragonal (i.e. elongated rock salt structure) on a SrTiO$_3$(100) substrate. 
In a subsequent theoretical study \cite{tet2} 
we investigated bulk phases of tetragonal CuO by
applying a hybrid density functional theory approach, which is able to deal
with the electronic, ionic and magnetic properties of such a system. Studying a variety of
magnetic orderings we obtained two energy minima as a function of tetragonal
distortion.  The antiferromagnetically ordered phase (denoted as TET2) with a tetragonal distortion of $c/a$=1.377 
appeared to be the most stable one with its lattice parameters being in very good agreement with
the experimental results. Furthermore, for the TET2 phase
we predicted a very high Ne{\'e}l temperature (T$_N$) of 800 K, which would perfectly
fit the trend of T$_N$ for the 3d-transition metal monoxides\cite{siemons}.
Also, the calculated density of states of the TET2 phase agreed well with the experimental 
valence band spectrum with the exception of a few residual differences related to the structure of the 
main peak and the presence of an additional peak in the low energy spectrum
which we have attributed to surface/substrate effects not taken into account in our previous analysis.

The aim of the present work is now, to corroborate our basic findings for the
artificial TET2 phase by modelling  experiment as realistically as
possible. This is what we do by describing the adsorbate system in terms of an 
$\rm (CuO)_{[n]}/SrTiO_3$(100) slab for varying thicknesses of CuO layers 
[with n denoting a CuO coverage of 2,3,3.5 and 4 unit cells (u.c.)]. 

Shortly after the publication of our bulk CuO study, a further theoretical
study  appeared, which was based on the self-interaction-corrected 
local-density functional method. \cite{peralta} Again, for a variety of
magnetic orderings two energy minima were found as a
function of the tetragonal distortion, but the $c/a$ values were much closer to 1 than in
our work and, consequently, in worse agreement with the experimentally observed $c/a$ ratio of 1.357.
Therefore, the present study also serves the purpose to buttress the
theoretical capability of our applied hybrid density functional theory
approach and to understand the nature of the exceptionally large vertical elongation observed in 
SrTiO$_3$ supported CuO.

\section{Computational Aspects}
Because standard density functional theory applications fail in correctly
describing the ground state of 3d-transition metal monoxides, a more sophisticated (and much
more costly)  approach has to be chosen.
Therefore, for the present study  we apply hybrid density functional
theory\cite{becke93} based on the Heyd-Scuseria-Ernzerhof (HSE)
method\cite{krukau} as implemented in the Vienna \emph{ab initio}
simulation package (VASP)\cite{gk1,gk2,gk3}. All the corresponding technical parameters
were the same as in our recent study \cite{tet2} of tetragonal bulk CuO. In
particular, 1/4 of short-ranged Hartree Fock exchange was admixed to the
generalized-gradient-approximation exchange-correlation functional. 
The adsorbate systems $\rm (CuO)_{[n]}/SrTiO_3$(100) (n=2,3,3.5 and
4 unit cells) were modelled by a repeated slab scheme containing up to 62 atoms
per slab for the highest coverage.  Figures \ref{fig:1} and \ref{fig:2} sketches the basic
layer-wise atomic arrangements.
All atomic positions of CuO layers were fully relaxed, whereas for the $\rm
SrTiO_3$ substrate the lowest 5 bottom layers, corresponding to a full $\rm SrTiO_3$ unit cell,
were kept fixed. 
The lateral lattice parameter $a$=3.90 \AA\, for $\rm SrTiO_3$ was
taken from the recent VASP-based HSE study by R. Wahl and coworkers \cite{wahl}.
Indeed, the optimal HSE value of $a$ is in very good agreement with the measured lattice
constant $a_{\rm Expt}$=3.900 \AA\, \cite{cao}.
In order to make the HSE computations feasible, ferromagnetic (FM) ordering was assumed although
a specific antiferromagnetic ordering appeared to be energetically more stable in
the ideal tetragonal bulk phase. \cite{tet2} Nevertheless, also because of the small magnetic Cu-moments
of about 0.7 $\mu_B$ we expect only a small influence of the magnetic
ordering on the structural relaxations and valence band spectrum\cite{franchinijcp}, 
which are the aim of study of our present work on the $\rm (CuO)_{[n]}/SrTiO_3$(100) substrate system.
The {\bf k}-points integration for the structural optimization runs has been done by using 
a 4$\times$4 two dimensional Monkhorst-Pack grid, which was increased to 6$\times$6 for
the final electronic relaxation.

\section{Results and Discussion}
\label{s3}

\subsection{Structural Properties} 

\begin{figure}
\includegraphics[clip,width=0.45\textwidth]{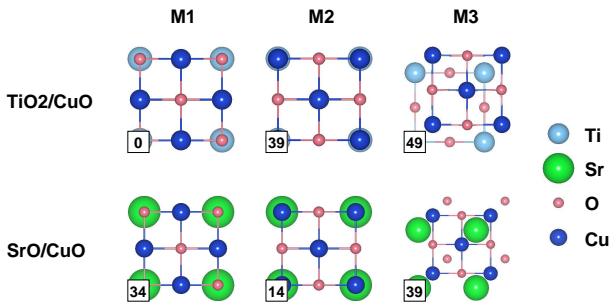}
\caption
{
(Color online) Top view of the  interface models of the $\rm
(CuO)_{[n]}/SrTiO_3$(100) adsorbate system as studied by the present HSE
approach.  The numbers in the squares (bottom left corner of each structure)
denote the calculated energy difference (in meV/\AA$^2$) relative to the TiO$_2$
terminated M1 stacking, which is the most stable arrangement.
}
\label{fig:1}
\end{figure}

For finding the energetically most stable atomic arrangement several
terminations of the $\rm SrTiO_3$(100) substrate and stacking of the CuO
adsorbate have been investigated.  Fig.  \ref{fig:1} sketches all the studied
cases: the substrate is either TiO$_2$ or SrO terminated. CuO layers may be
accommodated in several ways: (i) Cu on top of oxygen atoms (M1), (ii) Cu on top of Ti or Sr atoms
(M2) or (iii) Cu and O in  hollow sites (M3).  These (FM) calculations have been
done by placing 2 CuO unit cells (5 CuO layers) on SrTiO$_3$ (6 unit cells).
Fig. \ref{fig:1}  shows that the $\rm TiO_2$\ terminated M1 structure is the
most stable one.
It should, however,  be noted that rather close in energy (only 14 meV/\AA$^2$ less stable)
is the SrO terminated M2 stacking.  

\begin{figure}
\includegraphics[clip,width=0.45\textwidth]{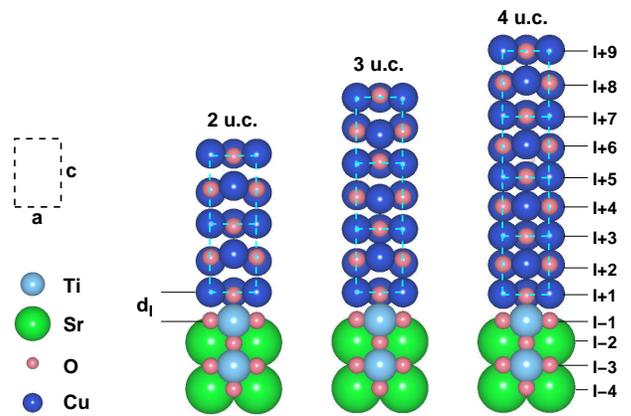}
\caption
{(Color online) Side view of the $\rm (CuO)_{[n]}/SrTiO_3$(100) slabs for a
coverage by $n$=2,3,4 unit cells of CuO as used in the HSE calculation. 
(The n=3.5 coverage, which was also studied, is not shown). Atomic
arrangements according to the TiO$_2$ terminated substrate and the M1 stacking
of the CuO adsorbate (see Fig. \ref{fig:1}).
The layers are numbered with respect to the interface: $\rm I+j$ denotes CuO layers, 
$\rm I-j$ marks $\rm SrTiO_3$ layers.  The side projected two dimensional unit 
cell with lattice parameters {\em a} and {\em c} is sketched by the dashed
rectangle.  Detailed structural data are listed in Table \ref{tab:1}. 
}
\label{fig:2}
\end{figure}

From now on, we only discuss the TiO$_2$
terminated M1 structure.  We study the evolution of the structural and
electronic properties of the $\rm (CuO)_{[n]}/SrTiO_3$(100) adsorbate system
for a CuO coverage of n=2,3,3.5 and 4 unit cells. This choice is based on the
experiments estimate, that a tetragonal CuO-layer consisting of 3 to 4 unit
cells can be grown on $\rm SrTiO_3$(100)\cite{siemons} .  
The corresponding structural models are displayed in Fig \ref{fig:2}.

%%%%%%%%%%%%%%%%%%%%%%%%%%%%%%%%%%%%%%%%%%%%%%%%%%%%
%%%%%%%%%%%%%%%%%%%%%%%%%%%%%%%%%%%%%%%%%%%%%%%%%%%%

\begin{figure}
\includegraphics[clip,width=0.45\textwidth]{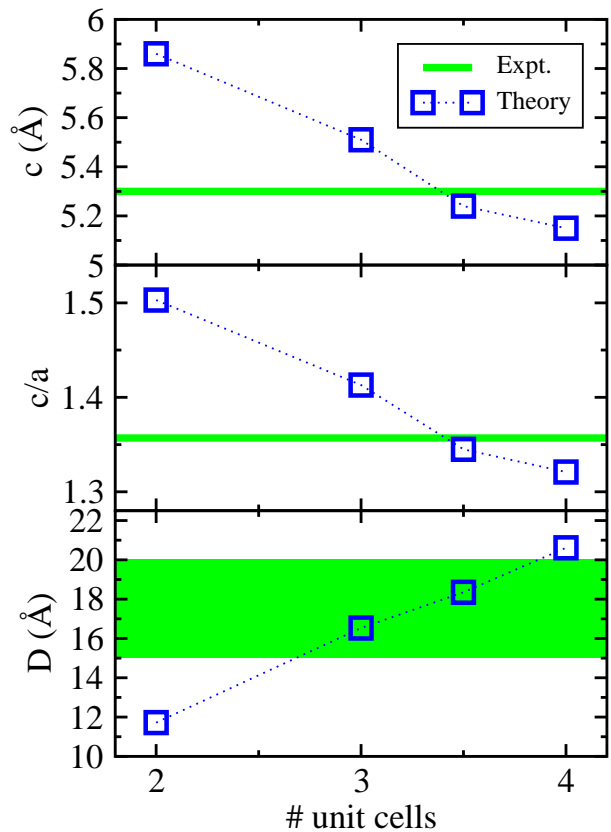}
\caption
{(Color online) Thickness dependent variation of the structural parameters
$\rm D$, $c/a$ and $c$ of the CuO adsorbate on $\rm SrTiO_3$(100)
as listed in Table \ref{tab:1}. Horizontal
bars illustrate the experimental data of Ref.\onlinecite{siemons}.
}
\label{fig:3}
\end{figure}

The structural properties are summarized in 
Fig.\ref{fig:3} and Table \ref{tab:1}.
Fig. \ref{fig:3} illustrates that theoretical and experimental geometrical
properties agree well when in the calculation the CuO coverage is 3.5 unit
cells, which is also in the estimated range of  the experimentally derived
thickness (parameter $\rm D$ in Table \ref{tab:1}).  The experimental
value of $c/a$=1.357 is close to the theoretical value of 1.345, which
compares favorably to $c/a$=1.377 as obtained from the HSE study for the
ideal bulk TET2 phase.\cite{tet2}  In general, when increasing the
number of unit cells $c$ and  $c/a$ decrease as shown in Table \ref{tab:1}.

The analysis of the interlayer distances reveals that inside the SrTiO$_3$
substrate the interface effects are healed out rather rapidly.  Only the
distance of the layer closest to the interface ($d$$\rm _{I-1/I-2}$) experiences a distinct
shortening, as it is also the case for the  CuO layer at the interface ($d$$\rm _{I+1/I-1}$). On the
other hand, the CuO layer distances appear to be quite sensitive to the
thickness of adsorbed CuO. For coverage of 2 and 3 unit cells the distances
are significantly larger than the reference value of 2.69 \AA\, which was
derived from the ideal TET2 bulk study. For coverage larger than 3 unit
cells the distances between the inner CuO layers are significantly shortened,
whereas the top 2 layers expand outwards.  This change in bond length is
reflected by the change of ionicity of Cu, as will be discussed in the
following.

%%%%%%%%%%%%%%%%%%%%%%%%%%%%%%%%%%%%%%%%%%%%%%%%%%%%
%%%%%%%%%%%%%%%%%%%%%%%%%%%%%%%%%%%%%%%%%%%%%%%%%%%%

%%%%%%%%%%%%%%%%%%%%%%%%%%%%%%%%%%%%%%%%%%%%%%%%%%%%
\begin{table}
\caption{
Geometrical data for $\rm (CuO)_{[n]}/SrTiO_3$(100) for n=2, 3, 3.5 and 4 (n:
number of CuO unit cells). The data for $c$ and $c/a$ are based on averages of
the vertical lattice parameter according to Ref. \onlinecite{siemons}, and the
lateral parameter $a$=3.90 \AA\, of SrTiO$_3$(100) is taken
from another VASP-based HSE study \cite{wahl}.  The total thickness of adsorbed CuO is
described by  $\rm D$.  The interlayer distance between two subsequent layers
as illustrated  in Fig.  \ref{fig:2} is denoted by $d$, and  $d$$\rm
_{I+1/I-1}$ indicates the interface separation between $\rm SrTiO_3$ and CuO.
Layers are labelled according to Fig.\ref{fig:2}.  The lattice parameters of
the HSE calculation for the ideal TET2 phase \cite{tet2} are $a$=3.908
and $c$=5.381 \AA\, resulting in an ideal bulk-like layer distance of 2.69
\AA. Experimental values are included when available. All length parameters
are given in units of \AA.
} \vspace{0.3cm}
\begin{ruledtabular}
\begin{tabular}{lccccc}
                             &  2 u.c. & 3 u.c. & 3.5 u.c & 4 u.c. & Expt.\cite{siemons} \\
\hline\hline
$c$                          &   5.86  &   5.51  & 5.24    & 5.15  & 5.3   \\
$c/a$                        &   1.503 &   1.413 & 1.345   & 1.321 & 1.357 \\
$\rm D$                &   11.72 &   16.53 & 18.35   & 20.61 & 15-20 \\
                             &        &         &         &       &\\
$d$$\rm _{I+9/I+8}$          &         &         &         & 2.82  &\\
$d$$\rm _{I+8/I+7}$          &         &         &  2.87   & 2.70  &\\
$d$$\rm _{I+7/I+6}$          &         &   2.84  &  2.80   & 2.63  &\\
$d$$\rm _{I+6/I+5}$          &         &   2.78  &  2.61   & 2.56  &\\
$d$$\rm _{I+5/I+4}$          &   3.03  &   2.85  &  2.57   & 2.53  &\\
$d$$\rm _{I+4/I+3}$          &   2.93  &   2.72  &  2.49   & 2.46  &\\
$d$$\rm _{I+3/I+2}$          &   2.90  &   2.65  &  2.44   & 2.43  &\\
$d$$\rm _{I+2/I+1}$          &   2.93  &   2.73  &  2.58   & 2.49  &\\
$d$$\rm _{I+1/I-1}$          &   2.47  &   2.37  &  2.37   & 2.29  &\\
                    &         &         &         &       &\\
$d$$\rm _{I-1/I-2}$          &   1.88  &   1.88  &  1.86   & 1.82  &\\
$d$$\rm _{I-2/I-3}$          &   2.00  &   2.00  &  1.97   & 1.92  &\\
$d$$\rm _{I-3/I-4}$          &   1.95  &   1.94  &  1.94   & 1.90  &\\
$d$$\rm _{I-4/I-5}$          &   1.97  &   1.97  &  1.94   & 1.90  &\\
$d$$\rm _{I-5/I-6}$          &   1.98  &   1.94  &  1.95   & 1.92  &\\
$d$$\rm _{I-6/I-7}$          &   1.98  &   1.95  &  1.95   & 1.92  &\\
$d$$\rm _{I-7/I-8}$          &  1.95  &   1.95  &  1.96   & 1.94  &\\
$d$$\rm _{bulk}$             &   1.95  &   1.95  & 1.95   &1.95  & 1.95 \\
\end{tabular}
\end{ruledtabular}
\label{tab:1}
\end{table}
%%%%%%%%%%%%%%%%%%%%%%%%%%%%%%%%%%%%%%%%%%%%%%%%%%%%

%%%%%%%%%%%%%%%%%%%%%%%%%%%%%%%%%%%%%%%%%%%%%%%%%%%%
%%%%%%%%%%%%%%%%%%%%%%%%%%%%%%%%%%%%%%%%%%%%%%%%%%%%

%%%%%%%%%%%%%%%%%%%%%%%%%%%%%%%%%%%%%%%%%%%%%%%%%%%%
\begin{figure}
\includegraphics[clip,width=0.45\textwidth]{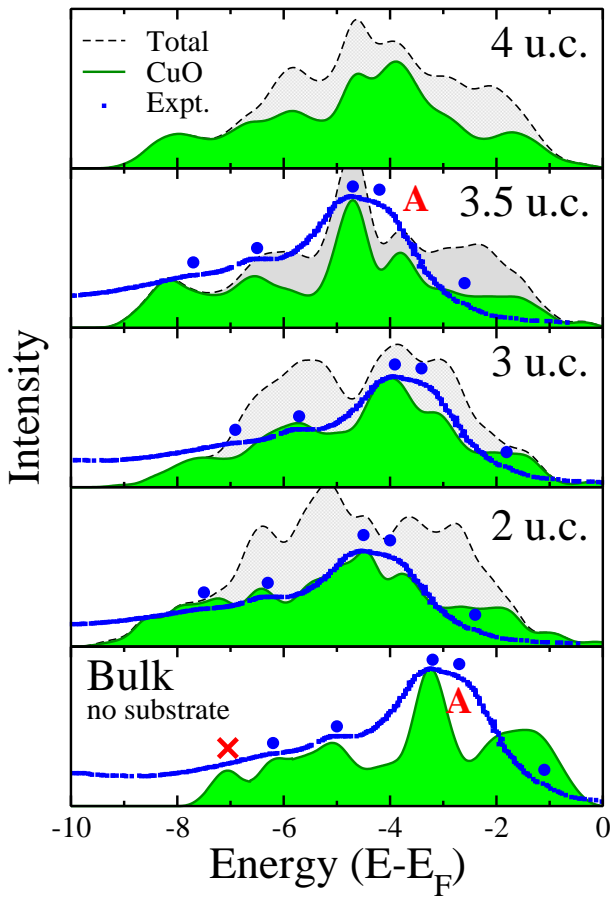}
\caption
{
(Color online) Results of the HSE calculations: total (light grey) and local
CuO-projected (dark green) density of states for (CuO)$\rm _{[n]}/\rm SrTiO_3$
[n=2, 3, 3.5 and 4 unit cells (u.c.)] and for the ideal bulk tetragonal TET2
phase.\cite{tet2}  The experimental curve is the valence-band spectrum of
Ref.\onlinecite{siemons}.  The circles and crosses indicate the position of the measured
peaks. Energy zero is chosen to be the top of the valence band. Position of
the maximum of the experimental spectrum is adjusted to the corresponding
maximum of the HSE density of states.  The calculated DOS is broadened by a Gaussian of
half-width 0.2 eV.
}
\label{fig:4}
\end{figure}
%%%%%%%%%%%%%%%%%%%%%%%%%%%%%%%%%%%%%%%%%%%%%%%%%%%%

\subsection{Electronic Properties}
Figure \ref{fig:4} compares the HSE calculated density of states (DOS) with
the experimental valence band spectrum. It is obvious that the total as
well as the local CuO-projected DOS is rather sensitive to the thickness of
the CuO block. Strikingly, the calculated DOS for the ideal TET2 bulk
phase and for the system with n=3-3.5 unit cells coverage agree well with
experiment, quite in contrast to all cases. 
Though the superstructure with 2 CuO u.c.
already displays some film features which are not reproduced in the calculated 
bulk spectrum (such as the appearance of the peak A), the overall comparison with the 
measured curve is not satisfactory due to (i) a much too dominant presence of substrate-related states
which weaken the intensity of the main CuO peaks around -4 eV and (ii) the lack of a
one-to-one correspondence between theoretical and experimental peaks.

As was discussed for the bulk phase TET2 \cite{tet2}, the agreement with experiment 
is rather good with the exception of one missing peak (indicated by the letter A in Fig. \ref{fig:4}) 
and one extra peak (indicated by the cross) in the calculated spectrum. 
From the DOS for n=3-3.5 one clearly deduces that the experimental structure
is reproduced very well (especially for 3.5 u.c.) and the overall 
agreement is improved with respect to the ideal bulk phase. 
In particular, a new peak (A) is found due to the CuO interface layer
(the local CuO DOS has a very distinctive peak at this energy position, which is also 
reflected in the total DOS) and the spurious peak found in the ideal bulk phase disappears.
Finally, according to the DOS for n=$3.5$ in Fig. \ref{fig:4} the highest experimental 
peak, can now be mainly attributed to the substrate, because there the local CuO DOS shows 
a weak depression, again in line with the observed valence band spectrum.

\begin{figure}
\includegraphics[clip,width=0.45\textwidth]{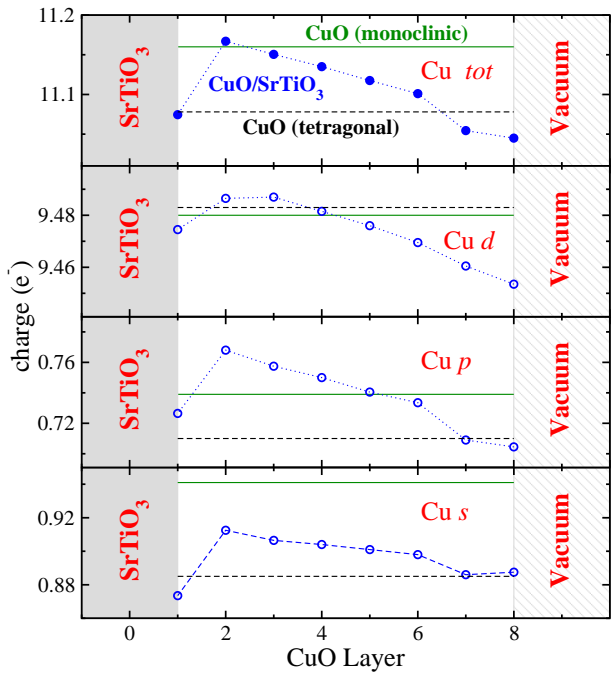}
\caption
{
(Color online) Layer by layer total and {\em l}-decomposed charge  of Cu atoms
(sphere radius of 1.16 \AA\,) of  (CuO)$\rm _{[3.5]}/\rm SrTiO_3$(100) in
comparison to the corresponding values for bulk monoclinic CuO (tenorite
structure) and the ideal bulk tetragonal TET2 phase.\cite{tet2} The
placements of the TiO$_3$ substrate and the vacuum regions are indicated.
}
\label{fig:5}
\end{figure}

Figure \ref{fig:5} illustrates the layer dependent change of ionicity of the
Cu atoms decomposed over the orbital quantum number $l$. A comparison with the 
corresponding bulk values for both the monoclinic and tenorite structures is also given.
Leaving out the layer distance at the interface there is a uniform
trend to be seen: with increasing distance from the interface the Cu charge is
reduced by about 0.15 $e^-$ (i.e. the ionicity of the positive Cu-ions increases). 
This change in charge/valency (topmost panel) is mostly due to the change of p- and d-like
charges (2nd and 3rd panel from the top). For the  two Cu layers closest to
the surface the total Cu charge is even less than for the tetragonal bulk
phase. This increase in valency is directly accompanied by an increase in the
interlayer distance, as clearly illustrated by Table \ref{tab:1}.

In general, an increase in bond length corresponds to an increase in ionicity:
typical Cu-O bond lengths  involving Cu$^{2+}$ ions are of the order of 2
\AA\, whereas for Cu$^{3+}$ the value might strongly increase up to 2.75
\AA\, (see Ref. \onlinecite{well87}). In our case, the increase
in interlayer distance  corresponds to the stretching of the Cu-O bond in the
vertical direction (perpendicular to the Cu-O layers). 
According to Table \ref{tab:1} for the topmost layer of the n=3.5 case the bond length is 2.87
\AA, being well in the bond length regime of  Cu$^{3+}$  ionicity. On the
other hand, the in-plane bond lengths --due to the SrTiO$_3$ substrate or the
lattice parameter $a$=3.91 \AA\, of the optimized bulk TET2 phase--
correspond rather well to an ionicity of Cu$^{2+}$. On the basis of our
present surface and recent bulk TET2 study  it seems that the Cu atoms
appear in two competing concurrent ionic states, depending on the directions of the Cu-O bonds.

Our interpretation of the layer-dependent modulation of the Cu electronic population and 
Cu-O bond lengths suggesting a peculiar coexistence of Cu$^{2+}$-like and Cu$^{3+}$-like
behavior is in line with the combined DFT and photoelectron spectroscopy study on copper
oxide clusters presented by L.-S. Wang and coworkers\cite{wang96}. By investigating the 
structural and electronic properties of Cu$2$O$_x$ ($x$=1-4) these authors report a 
significant change both in the Cu-O bond length (0.05 \AA) and Cu charge (0.1$e^-$) going from 
Cu$_2$O$_2$ (Cu$^{2+}$) to Cu$_2$O$_3$ (Cu$^{3+}$).

We can attribute the relatively small charge difference between bulk and surface Cu ions
(0.15 $e^-$), to the aforementioned completion between Cu$^{2+}$-like and Cu$^{3+}$-like environments.
To support further this interpretation we have calculated the charge difference between
2+ and 3+ Mn ions in MnO and Mn$_2$O$_3$ using the optimized data provided in Ref. \onlinecite{mno}.
Indeed, the resulting value of 0.38 $e^-$, by far smaller than the expected difference of 1 electron 
which would result from a simplified ionic picture, is twice larger than the corresponding Cu-difference 
in our frustrated CuO/SrTiO$_3$ superstructure.

Finalizing the discussion of the electronic structure, the gap of the ideal 
TET2 phase of 2.7 eV is strongly reduced when CuO is grown on $\rm SrTiO_3$:
for a coverage of 3.5 unit cells the gap formed by CuO states is reduced to
about 0.6 eV, which for the total system (now including substrate states) is
even more reduced to about 0.5 eV. The average local magnetic moment for Cu is
about 0.75 $\mu_B$, which is enhanced in comparison to 0.63 $\mu_B$ of the bulk
TET2 phase. It should be noted that the present calculation for the surface
system $\rm (CuO)_{[n]}/SrTiO_3$(100) was done for ferromagnetic ordering (in
order to make the HSE calculation feasible) whereas for
the bulk TET2 calculation the most stable phase was found for some particular
antiferromagnetic ordering \cite{tet2}. However, in
Ref. \onlinecite{tet2} studying several magnetic orderings it was found, that 
the local Cu moment is rather insensitive to the specific  alignment of spins. 

%%%%%%%%%%%%%%%%%%%%%%%%%%%%%%%%%%%%%%%%%%%%%%%%%%%%
%%%%%%%%%%%%%%%%%%%%%%%%%%%%%%%%%%%%%%%%%%%%%%%%%%%%

\subsection{Summary}
Summarizing, our first principles study on the $\rm (CuO)_{[n]}/SrTiO_3$(100)
adsorbate system based on a hybrid density functional theory approach
describes and explains the recent experiments \cite{siemons} rather
well, which corroborates our findings and predictions for the ideal bulk (without 
$\rm SrTiO_3$ substrate) TET2 phase \cite{tet2}.  The analysis of our calculated results for structural and
electronic properties indicates that the physical properties of the Cu-O bonds are rather peculiar.
In particular, we find that the enormous structural anisotropy of the Cu-O
sublattice --which determines the experimentally observed tetragonal
symmetry-- can be understood in terms of a layer dependent evolution of the
Cu ionicity which increases progressively towards the surface.
Ultimately, the local structural and electronic Cu-O environment 
appears very frustrated as a results of the coexistence between two
concurrent states attributable to an in-plane Cu$^{2+}$-like and out-of-plane Cu$^{3+}$-like
arrangements.

\section{Acknowledgments}
%%%%%%%%%%%%%%%%%%%%%%%%%%% Acknowledgment %%%%%%%%%%%%%%%%%%%%%%%%%%%%%%%%%
Research in Vienna was sponsored by the FP7 European Community grant ATHENA.
Support by the FWF, project nr. F4110-N13 is gratefully acknowledged. 
Research at the Shenyang National Laboratory for Materials Science
was sponsored by the Materials Processing Modeling Division. X.-Q.C
acknowledges the support from the ‘‘Hundred Talents
Project’’ of Chinese Academy of Science.
All calculations have been performed on the Vienna Scientific Cluster (VSC).
%%%%%%%%%%%%%%%%%%%%%%%%%%%%%%%%%%%%%%%%%%%%%%%%%%%%%%%%%%%%%%%%%%%%%%%%%%%%

\end{document}